\documentclass[prd,showkeys,twocolumn,showpacs,preprintnumbers,amsmath,amssymb]{revtex4}
\usepackage{graphicx}% Include figure files
\usepackage{dcolumn}% Align table columns on decimal point
\usepackage{bm}% bold math

\renewcommand{\d}{\mathrm{d}}
\renewcommand{\Re}{\mathrm{Re\:}}
\renewcommand{\Im}{\mathrm{Im\:}}
\newcommand{\e}{\mathrm{e}}

\begin{document}

\title{Derivative dispersion relations above the physical threshold}

\author{R.F. \'Avila}
\affiliation{%
Instituto de Matem\'atica, Estat\'{\i}stica e
Computa\c c\~ao Cient\'{\i}fica\\
Universidade Estadual de Campinas, UNICAMP\\
13083-970 Campinas, SP, Brazil}
\author{M.J. Menon}
\email{menon@ifi.unicamp.br}
\affiliation{%
Instituto de F\'{\i}sica Gleb Wataghin\\
Universidade Estadual de Campinas, UNICAMP\\
13083-970 Campinas, SP, Brazil}

\date{\today}% It is always \today, today,

\begin{abstract}
We discuss some formal and practical aspects related to the 
replacement of Integral Dispersion Relations (IDR) by derivative 
forms, \textit{without high-energy approximations}. We first 
demonstrate that, for a class of functions with physical interest
as forward scattering amplitudes, this replacement can be analytically
performed, leading to novel Extended Derivative Dispersion Relations
(EDDR), which, in principle, are valid for any energy above the
physical threshold. We then verify the equivalence between the IDR and
EDDR by means of a popular parametrization for total cross sections 
from proton-proton and antiproton-proton scattering and compare the 
results with those obtained through other representations for the 
derivative relations. Critical aspects on the limitations of the whole
analysis, from both formal and practical points of view, are also 
discussed in some detail.
\end{abstract}



\vspace{0.5cm}

\centerline{Published in \textit{Brazilian Journal of Physics} \textbf{37}, 358
(2007)}

\vspace{0.5cm}

\maketitle

\section{Introduction}

Elastic hadron scattering constitutes a hard challenge for QCD. The
problem concerns the large distances involved (confinement), which
renders difficult the development of a formal nonperturbative
calculational scheme for scattering states, able to describe soft
diffractive processes. At this stage Analyticity, Unitarity, Crossing
and their consequences, still represent a fundamental framework for
the development of theoretical ideas, aimed to reach efficient
descriptions of the experimental data involved. In this context,
dispersion relations, connecting real and imaginary parts of the
scattering amplitude, play an important role as a useful mathematical
tool, in the simultaneous investigation of particle-particle and
antiparticle-particle scattering.

Dispersion relations in integral form, for hadronic amplitudes, were
introduced in the sixties, as consequences of the Cauchy's theorem and
the analytic properties of the scattering amplitude, dictated by 
unitarity \cite{idr1,idr2,bc}. However, two kinds of limitations
characterize this \textit{integral} approach: (1) its nonlocal 
character (in order to evaluate the real part, the imaginary part must
be known in all the integration space); (2) the restricted class of 
functions that allows analytical integration. Later on it was shown
that, for hadronic forward elastic scattering 
\textit{in the region of high and asymptotic energies}, these integral
relations can be replaced by derivative forms \cite{fddr,bks,kn}. 
Since then, the formal replacement of integral by derivative relations
and their practical use have been widely discussed in the literature
\cite{ed,hk,hjk,bd,ddr-others,kf,kf-blois}, mainly in the seminal 
papers by Kol\'a\v{r} and Fischer \cite{kf,kf-blois}. See  
Ref. \cite{am04} for a recent critical review on the subject.

Despite the  results that have been obtained with the 
\textit{derivative} approach, the high-energy condition (specifically,
center-of-mass energies above 10 - 20 GeV) turns out difficult any
attempt to perform global fits to the experimental data connecting 
information from low and high energy regions. A first step in this 
direction appears in Ref. \cite{cms}, where  new representations for
the derivative relations, extended to low energies, have been
introduced by Cudell, Martynov and Selyugin and to which we shall 
refer in what follows. However, a rigorous formal extension of the 
derivative dispersion relations down to the physical threshold,
providing a complete analytical equivalence between integral and 
differential approaches, is still missing and that is the point we are
interested in.

In this work, we first demonstrate that, for a class of functions of 
physical interest as forward elastic scattering amplitudes, the 
integral relations can be analytically replaced by derivative forms
without the high-energy approximation. Therefore, in principle, for
this class of functions, derivative relations hold for any energy 
above the physical threshold. We then check the consistences of the
results obtained with the integral relations and the extended 
derivative dispersion relations  by means of a simple analytical 
parametrization for the total cross sections from proton-proton ($pp$)
and antiproton-proton ($\bar{p}p$) scattering (highest energy interval
with available data). In addition, we compare the results with those 
obtained through the standard derivative relations (high-energy 
condition) and the derivative representation by 
Cudell-Matynov-Selyugin. We shall show that, above the physical 
threshold, only the extended relations lead to exactly the same 
results as those obtained with integral forms. We proceed with a 
critical discussion on the limitations of our analysis from both 
formal and practical points of view.

The manuscript is organized as follows. In Sec. \ref{sec:DR} we recall
the main formulas and some conditions involving the 
\textit{Integral Dispersion Relations} (IDR), the \textit{standard 
Derivative Dispersion Relations} (sDDR) and the 
\textit{Cudell-Martynov-Selyugin representations} (CMSr); we also
present, in certain detail, the replacement of IDR by the 
\textit{Extended Derivative Dispersion Relations} (EDDR). In Sec. 
\ref{sec:pra} we check the consistences and exemplify the 
applicability of all these results in simultaneous fits to the total 
cross section and the ratio $\rho$ of the real to imaginary parts of 
the forward amplitude, from $pp$ and $\bar{p}p$ scattering. In Sec.
\ref{sec:critical} we present a critical discussion on all the 
obtained results. The conclusions and some final remarks are the 
contents of Sec. \ref{sec:conclusions}.

\section{Dispersion Relations \label{sec:DR}}

\subsection{Integral Dispersion Relations \textrm{(IDR)}}

First, it is important to recall that Analyticity, Unitarity and 
Crossing lead to IDR for the scattering amplitudes in terms of a 
\textit{crossing symmetric variable}. For an elastic process, 
$m + m \rightarrow m + m$, in the forward direction, this variable 
corresponds to the energy of the incident particle in the laboratory
system, $E$ \cite{idr1}. In this context and taking into account
polynomial boundedness, the one subtracted IDR for crossing even ($+$)
and odd ($-$) amplitudes, in the physical region 
($E:m \rightarrow \infty$), read \cite{idr1,idr2,bc}

\begin{equation}
\Re F_{+}(E)= K + \frac{2E^{2}}{\pi}P\!\!\!\int_{m}^{+\infty}
\!\!\!\d E'
\frac{1}{E'(E'^{2}-E^{2})}\Im F_{+}(E'),
\label{eq:idre}
\end{equation}
\begin{equation}
\Re F_{-}(E)=  \frac{2E}{\pi}P\!\!\!\int_{m}^{+\infty} \!\!\!
\d E'
\frac{1}{(E'^{2}-E^{2})}\Im F_{-}(E'),
\label{eq:idro}
\end{equation}
where $K$ is the subtraction constant.

The connections with the hadronic amplitudes for crossed channels,
such as $pp$ and $\bar{p}p$ elastic scattering,
are given by the usual definitions:

\begin{equation}
F_{pp} = F_{+} + F_{-} \qquad
F_{\bar{p}p} = F_{+} - F_{-}.
\label{eq:3}
\end{equation}

The main practical use of the IDR concerns simultaneous investigations
on the total cross section (Optical Theorem) and the ratio $\rho$ of 
the real to imaginary parts of the forward amplitude, which is also 
our interest here. In terms of the crossing symmetric variable $E$ 
these physical quantities are given, respectively, by \cite{bc}

 \begin{equation}
\sigma_{\mathrm{tot}}
=\frac{4\pi}{\sqrt{E^2-m^2}} \Im F(E,\theta_{\mathrm{lab}}=0),
\label{eq:tcs}
\end{equation}

\begin{equation}
\rho(E) = \frac{\Re F(E,\theta_{\mathrm{lab}}=0)}
{\Im F(E,\theta_{\mathrm{lab}}=0)},
\label{eq:rho}
\end{equation}
where $\theta_{\mathrm{lab}}$ is the scattering angle in the 
laboratory system.

\subsection{Standard Derivative Dispersion Relations 
\textrm{(sDDR)}\label{sec:SDDR}}

Basically, at high energies, the replacement of IDR by sDDR is 
analytically performed by considering the limit $m \rightarrow 0$ in
Eqs. (\ref{eq:idre}) and (\ref{eq:idro}) \cite{bks,am04}. It should be
recalled that an additional high-energy approximation is considered in
these integral equations, when they are expressed in terms of the 
center-of-mass energy squared $s =2(m^2 + mE)$ and not $E$ \cite{bks}.
However, based on a rigorous replacement (discussed in 
Sec. \ref{sec:EDDR}), we consider the derivative relations in terms of
the crossing symmetric variable $E$. In this case the sDDR read  
\cite{bks,kf,am04}

\begin{equation}
\Re F_{+}(E)= K +
E\tan\left[\frac{\pi}{2}\frac{\mathrm{d}}{\mathrm{d}\ln E} \right]
\frac{\Im F_{+}(E)}{E},
\label{eq:sddre}
\end{equation}

\begin{equation}
\Re F_{-}(E)
=\tan\left[\frac{\pi}{2}
\frac{\mathrm{d}}{\mathrm{d}\ln E} \right]
\Im F_{-}(E).
\label{eq:sddro}
\end{equation}

Necessary and sufficient conditions for the convergence of the
above tangent series have been established by Kol\'a\v{r} and Fischer,
in particular through the following theorem \cite{kf}:

\newtheorem{guess}{Theorem \label{theo:1}}
\begin{guess}
Let $f: R^1 \rightarrow R^1$. The series

\begin{eqnarray}
\tan\left[\frac{\pi}{2} \frac{\mathrm{d}}{\mathrm{d}x} \right] f(x)
\nonumber
\end{eqnarray}
converges at a point $x \in R^1$ if and only if the series
\begin{eqnarray}
\sum_{n=o}^{\infty} f^{(2n + 1)}(x)
\nonumber
\end{eqnarray}
is convergent.
\end{guess}

For example, in the case of $f(x) = e^{\gamma x}$, $\gamma$ a real 
constant, the ratio test demands $|\gamma| < 1$ for the series to be 
absolutely convergent (which will be our interest in 
Sec. \ref{sec:pra}).

\subsection{Cudell-Martynov-Selyugin Representations 
\textrm{(CMSr)}\label{sec:CMSR}}

Recently, the following representations have been introduced for the
derivative dispersion relations \cite{cms}

\begin{eqnarray}
\Re F_{+}(E)&=& K +
E\tan\left[\frac{\pi}{2}\frac{\mathrm{d}}{\mathrm{d}\ln E} \right]
\frac{\Im F_{+}(E)}{E} \nonumber \\
&&-
\frac{2}{\pi}
\sum_{p=0}^{\infty}
\frac{C_+(p)}{2p+1}\left(\frac{m}{E}\right)^{2p},
\label{eq:cmsre}
\end{eqnarray}
where,

\begin{displaymath}
C_+(p)=\frac{\e^{-\xi D_\xi}}{2p+1+D_\xi}
\left[\Im F_+(E)-E\Im F'_+(E)\right].
\end{displaymath}
and

\begin{eqnarray}
\Re F_{-}(E)&=&
-E\cot\left[\frac{\pi}{2}\frac{\mathrm{d}}{\mathrm{d}\ln E} \right]
\frac{\Im F_{-}(E)}{E} \nonumber \\
&-&
\frac{2}{\pi}
\sum_{p=0}^{\infty}
\frac{C_-(p)}{2p+1}\left(\frac{m}{E}\right)^{2p+1},
\label{eq:cmsro}
\end{eqnarray}
where

\begin{displaymath}
C_-(p)=\frac{\e^{-\xi D_\xi}}{2p+D_\xi}
\left[\Im F'_+(E)\right],
\end{displaymath}
and $\xi=\ln(E/m)$ and
$D_\xi=\frac{\mathrm{d}}{\mathrm{d}\xi}$.

We note the presence of correction terms in the form of infinity
series, which go to zero as the energy increases, leading to the
sDDR, Eqs. (\ref{eq:sddre}-\ref{eq:sddro}).
We shall use this representation in Sec.~\ref{sec:pra}, where
their applicability is discussed in detail.

\subsection{Extended Derivative Dispersion Relations 
\textrm{(EDDR)}\label{sec:EDDR}}

In this section we present our analytical replacement of the IDR by 
derivative forms without the high-energy approximation. We also 
specify the class of functions for which this replacement can be 
formally performed.

Let us consider the even amplitude, Eq. (\ref{eq:idre}). Integrating
by parts we obtain

\begin{eqnarray}
\label{eq:10}
\Re F_+(E)&=&
K - \frac{E}{\pi}\ln\left|\frac{m-E}{m+E}\right|\frac{\Im F_+(m)}{m} \\
&&-\frac{E}{\pi}\int_{m}^{\infty}
{\ln\left|\frac{E'-E}{E'+E}\right|
\frac{\d}{\d E'}\frac{\Im F_+(E')}{E'}\d E'}.
\nonumber
\end{eqnarray}

Following Ref. \cite{cms}, we define $E'=m \e^{\xi'}$ and 
$E=m \e^{\xi}$, so that the integral term in the above formula is 
expressed by

\begin{equation}
\label{eq:11}
\frac{m\e^\xi}{\pi} \int_{m}^{\infty}
\ln \coth\left(\frac{1}{2}|\xi'-\xi|\right)
\frac{\d}{\d \xi'}g(\xi')\d \xi',
\end{equation}
where $g(\xi')=\Im F(m\e^{\xi'})/(m\e^{\xi'})$.
Expanding the logarithm in the integrand in powers of $x=\xi'-\xi$,
\begin{displaymath}
\ln \left(\cot \frac{1}{2}|x|\right)
=\ln\left(\frac{1+\e^{-|x|}}{1-e^{-|x|}}\right)
=2\sum_{p=0}^{\infty}\frac{\e^{-(2p+1)|x|}}{2p+1},
\end{displaymath}
and assuming that

\begin{eqnarray}
\frac{\d}{\d\xi'}g(\xi') \equiv \tilde{g}(\xi')
\nonumber
\end{eqnarray}
is an analytic function of its argument,
we perform the expansion
\begin{eqnarray}
\tilde{g}(\xi')&=&
\sum_{n=0}^{\infty}\left.
\frac{{\mathrm{d}}^{n}}{{\mathrm{d}}\xi'^{n}}\tilde{g}(\xi')
\right|_{\xi'=\xi}\frac{(\xi'-\xi)^n}{n!}
\nonumber \\
& = & \sum_{n=0}^{\infty}\frac{\tilde{g}^{(n)}(\xi)}{n!}(\xi'-\xi)^n.
\nonumber
\end{eqnarray}

Substituting the above formulas in Eq.~(\ref{eq:11}) and integrating 
term by term, \textit{under the assumption of uniform convergence of 
the $\tilde{g}(\xi')$ series}, we obtain 

\begin{displaymath}
\frac{2m\e^\xi}{\pi}
\sum_{p=0}^{\infty}\frac{1}{2p+1}
\sum_{k=0}^{\infty}
\frac{1}{k!}\frac{\d^{k}}{\d \xi^k}\tilde{g}(\xi)
I_{kp},
\end{displaymath}
where
\begin{widetext}
\begin{eqnarray}
I_{kp}=\int_{0}^{\infty}\e^{-(2p+1)|\xi'-\xi|}(\xi'-\xi)^k\d \xi'
=
\frac{1}{(2p+1)^{k+1}}[((-1)^k+1)k!-(-1)^k\Gamma(k+1,(2p+1)\xi)]
\nonumber
\end{eqnarray}
\end{widetext}
and $\Gamma$ is the incomplete gamma function
$\Gamma(a,z)=\int_{z}^\infty t^{a-1}\e^{-t}\ \mathrm{d} t$.

With this procedure and from $\xi=\ln (E/m)$, Eq.~(\ref{eq:10}) is 
expressed by
\begin{widetext}
\begin{eqnarray}
\Re F_+(E)&=&
K-
\frac{E}{\pi}\ln\left|\frac{m-E}{m+E}\right|\frac{\Im F_+(m)}{m}
+\frac{4E}{\pi}\sum_{p=0}^{\infty}\sum_{k=0}^{\infty}
\frac{1}{(2p+1)^{2k+2}}
\frac{\d^{2k+1}}{\d(\ln E)^{2k+1}}\frac{\Im F_{+}(E)}{E}
\nonumber \\
&&+\frac{2E}{\pi}
\sum_{k=0}^{\infty}\sum_{p=0}^{\infty}
\frac{(-1)^{k+1}\Gamma(k+1,(2p+1)\xi)}{(2p+1)^{k+2}k!}
\frac{\d^{k+1}}{\d (\ln E)^{k+1}}
\frac{\Im F_+(E)}{E}, \nonumber
\end{eqnarray}
which can be put in the final form

\begin{equation}
\Re F_{+}(E)=
K+E\tan\left(\frac{\pi}{2}\frac{{\mathrm{d}}}{{\mathrm{d}}\ln
E}\right)\frac{\Im F_+(E)}{E}
+\Delta^+(E,m),
\label{eq:eddre}
\end{equation}

where the correction term $\Delta^{+}$ is given by
\begin{eqnarray}
\Delta^{+}(E,m)=
-\frac{E}{\pi}\ln\left|\frac{m-E}{m+E}\right|\frac{\Im F_+(m)}{m}
+\frac{2E}{\pi}
\sum_{k=0}^{\infty}\sum_{p=0}^{\infty}
\frac{(-1)^{k+1}\Gamma(k+1,(2p+1)\ln(E/m))}{(2p+1)^{k+2}k!}
\frac{\d^{k+1}}{\d (\ln E)^{k+1}}
\frac{\Im F_+(E)}{E}.
\nonumber
\end{eqnarray}

With analogous procedure for the odd relation we obtain

\begin{equation}
\Re F_{-}(E)=
\tan\left(\frac{\pi}{2}
\frac{{\mathrm{d}}}{{\mathrm{d}}\ln E}\right)\Im F_-(E)
+\Delta^-(E,m),
\label{eq:eddro}
\end{equation}
where

\begin{eqnarray}
\Delta^-(E,m)=
-\frac{1}{\pi}\ln\left|\frac{m-E}{m+E}\right|\Im F_-(m)
+\frac{2}{\pi}
\sum_{k=0}^{\infty}\sum_{p=0}^{\infty}
\frac{(-1)^{k+1}\Gamma(k+1,(2p+1)\ln(E/m))}{(2p+1)^{k+2}k!}
\frac{\d^{k+1}}{\d (\ln E)^{k+1}}
\Im F_-(E).
\nonumber
\end{eqnarray}
\end{widetext}

Equations (\ref{eq:eddre}) and (\ref{eq:eddro}) are the novel
EDDR, which are valid, in principle, for any energy
\textit{above} the physical threshold, $E = m$.

We note that the
correction terms $\Delta ^\pm \rightarrow 0$ as $E \rightarrow 
\infty$, leading, in this case, to the sDDR, Eqs. (\ref{eq:sddre})
and (\ref{eq:sddro}). We also note that the structure of the CMSr, 
Eqs. (\ref{eq:cmsre}) and (\ref{eq:cmsro}), are similar to the above
results, but without the logarithm terms. These terms come from the 
evaluation of the primitive at the lower limit in the integration by 
parts.

Since Theorem \ref{theo:1} insures the uniform convergence of the 
series expansion associated with
\begin{equation}
\tilde{g} = \frac{\mathrm{d}}{\mathrm{d}\ln E} \frac{\Im F(E)}{E}
\end{equation}
the condition imposed by this Theorem defines the class of functions 
for which the EDDR hold. For example, that is the case for 
$f(x) = e^{\gamma x}$, $0 < \gamma < 1$, referred to in 
Sec. \ref{sec:SDDR}. Other conditions are discussed by Kol\'a\v{r} and
Fischer \cite{kf}.

\section{Practical Equivalences between the Integral and
the Derivative Approaches} \label{sec:pra}

In this section we verify and discuss the consistences between the 
analytical structures of the IDR and the EDDR in a specific example: 
the connections of the total cross section with the $\rho$ parameter 
from $pp$ and $\bar{p}p$ scattering. Firstly, it is important to note
that the efficiency of both integral and derivative approaches in the 
description of the experimental data, depends, of course, on the 
theory available, namely the input for the imaginary part of the 
amplitude. In the absence of a complete model, valid for any energy 
above the physical threshold, we shall consider only as a 
\textit{framework}, a Pomeron-Reggeon parametrization for the 
scattering amplitude \cite{cmg,ckk}. For $pp$ and $\bar{p}p$ 
scattering this analytical model assumes nondegenerate contributions
from the even ($+$) and odd ($-$) secondary reggeons 
($a_2$/$f_2$ and $\rho$/$\omega$, respectively), together with a 
simple pole Pomeron contribution:

\begin{equation}
\Im F(E)=
XE^{\alpha_{\tt I\!P}(0)}+Y_+E^{\alpha_+(0)}+\tau Y_-E^{\alpha_-(0)},
\label{eq:15}
\end{equation}
where $\tau=+1$ for $pp$ and $\tau=-1$ for $\bar{p}p$. As usual, the 
Pomeron and the even/odd reggeon intercepts are expressed by

\begin{equation}
\alpha_{\tt I\!P}(0)=1+\epsilon, \qquad
\alpha_{+/-}(0)=1-\eta_{+/-}.
\label{eq:16}
\end{equation}

We stress that the Pomeron-Reggeon phenomenology is intended for the 
high-energy limit (rigorously, $E$ or $\sqrt s \rightarrow \infty$). 
Its use here, including the region of low energies, has only a 
framework character. However, as we shall show, this model is 
sufficient for a comparative analysis of the consistences. We shall 
return to this aspect in Sec. \ref{sec:critical}.

In what follows, the point is to treat simultaneous fits to the total 
cross section and the $\rho$ parameter from $pp$ and $\bar{p}p$ 
scattering and compare the results obtained with both IDR and EDDR.
Schematically, with parametrization (\ref{eq:15}-\ref{eq:16}) we 
determine $\Im F_{+/-}(E)$ through Eq. (\ref{eq:3}) and then 
$\Re F_{+/-}(E)$ either by means of the IDR,
Eqs. (\ref{eq:idre}-\ref{eq:idro}) or the EDDR,
Eqs. (\ref{eq:eddre}-\ref{eq:eddro}). Returning to Eq. (\ref{eq:3}) we
obtain $\Re F_{pp}(E)$ and $\Re F_{\bar{p}p}(E)$ and, at last,  
Eqs. (\ref{eq:tcs}) and (\ref{eq:rho}) lead to the analytical 
connections between $\sigma_{\mathrm{tot}}(E)$ and $\rho(E)$ for both 
reactions. Moreover, through the same procedure, we shall also compare
the above results with those obtained by means of both the sDDR,
Eqs. (\ref{eq:sddre}-\ref{eq:sddro}) and the CMSr,
Eqs. (\ref{eq:cmsre}-\ref{eq:cmsro}). We first present the fit 
procedure and then discuss all the obtained results.

\subsection{Fitting and Results \label{sec:fit}}

For the experimental  data on $\sigma_{\mathrm{tot}}(s)$ and 
$\rho(s)$, we made use of the Particle Data Group archives \cite{pdg},
to which we added the values of $\rho$ and $\sigma_{\mathrm{tot}}$ 
from $\bar{p}p$ scattering at 1.8 TeV, obtained by the E811 
Collaboration \cite{cavila}. The statistical and systematic errors 
were added in quadrature. The fits were performed through the 
CERN-Minuit code, with the estimated errors in the free parameters 
corresponding to an increase of the $\chi^2$ by one unit. To fit the 
data as function of the center-of-mass energy, we express the lab 
energy in the corresponding formulas in terms of $s$, namely 
$E = (s - 2m^2)/2m$.

We included all the data above the physical threshold, $\sqrt s > 2m
\approx 1.88$ GeV, that is, we did not perform any kind of data 
selection. Since the ensemble has a relatively large number of 
experimental points just above the threshold, the statistical quality
of the fit is limited by the model used here as framework. In fact,
with the these choices and procedures we obtained reasonable 
statistical results (in terms of the $\chi^2$ per degree of freedom)
only for an energy cutoff of the fits at  $\sqrt s_{\mathrm{min}} =$ 
4 GeV. However, we stress that our focus here is in tests on the 
consistences among the different relations and representations and 
not, strictly, on the statistical quality of the fits (we shall return
to this point in Sec. \ref{sec:critical}).

\begin{figure}[!]
\includegraphics[width=8.cm,height=6.cm]{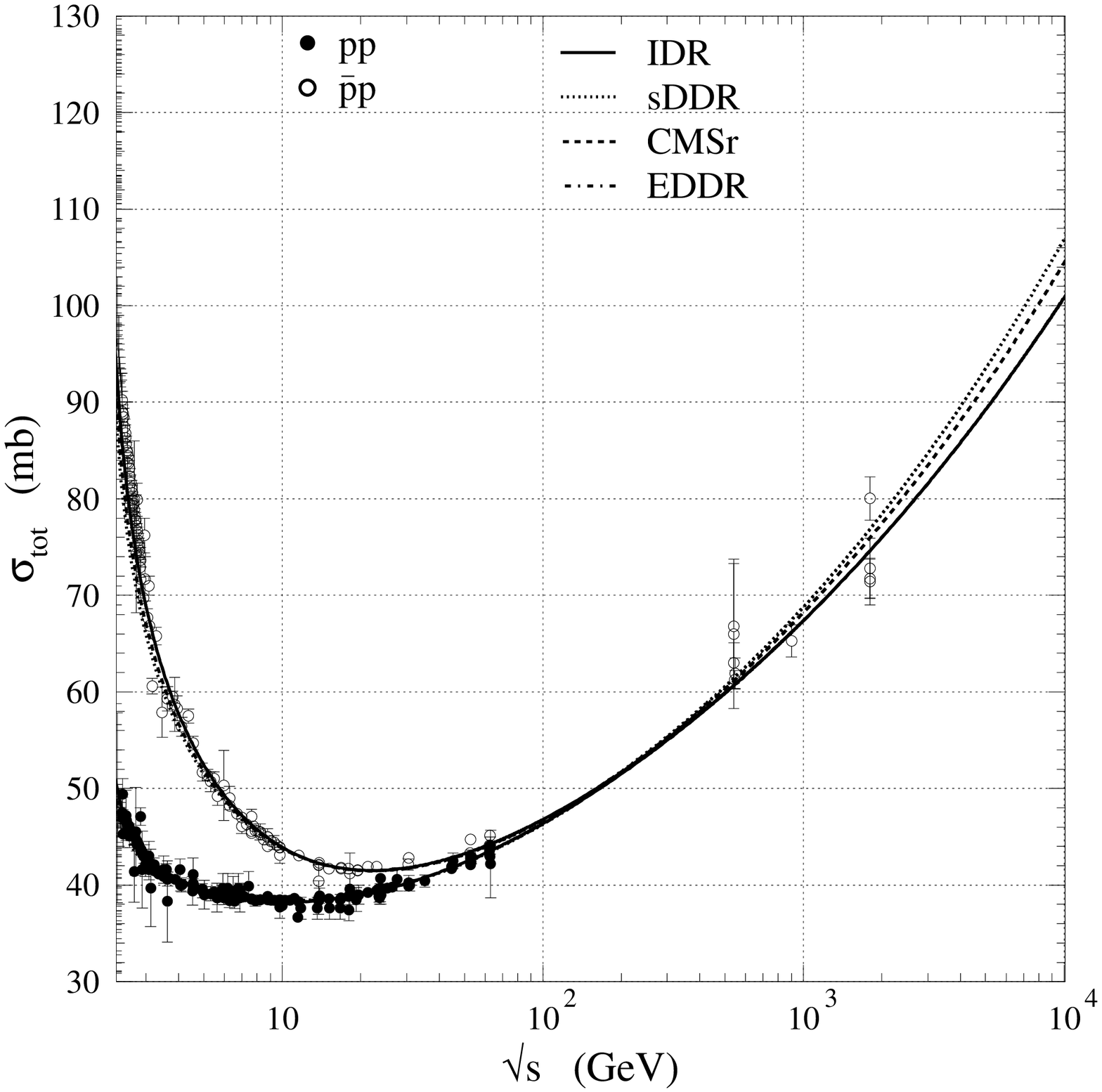}
\includegraphics[width=8.cm,height=6.cm]{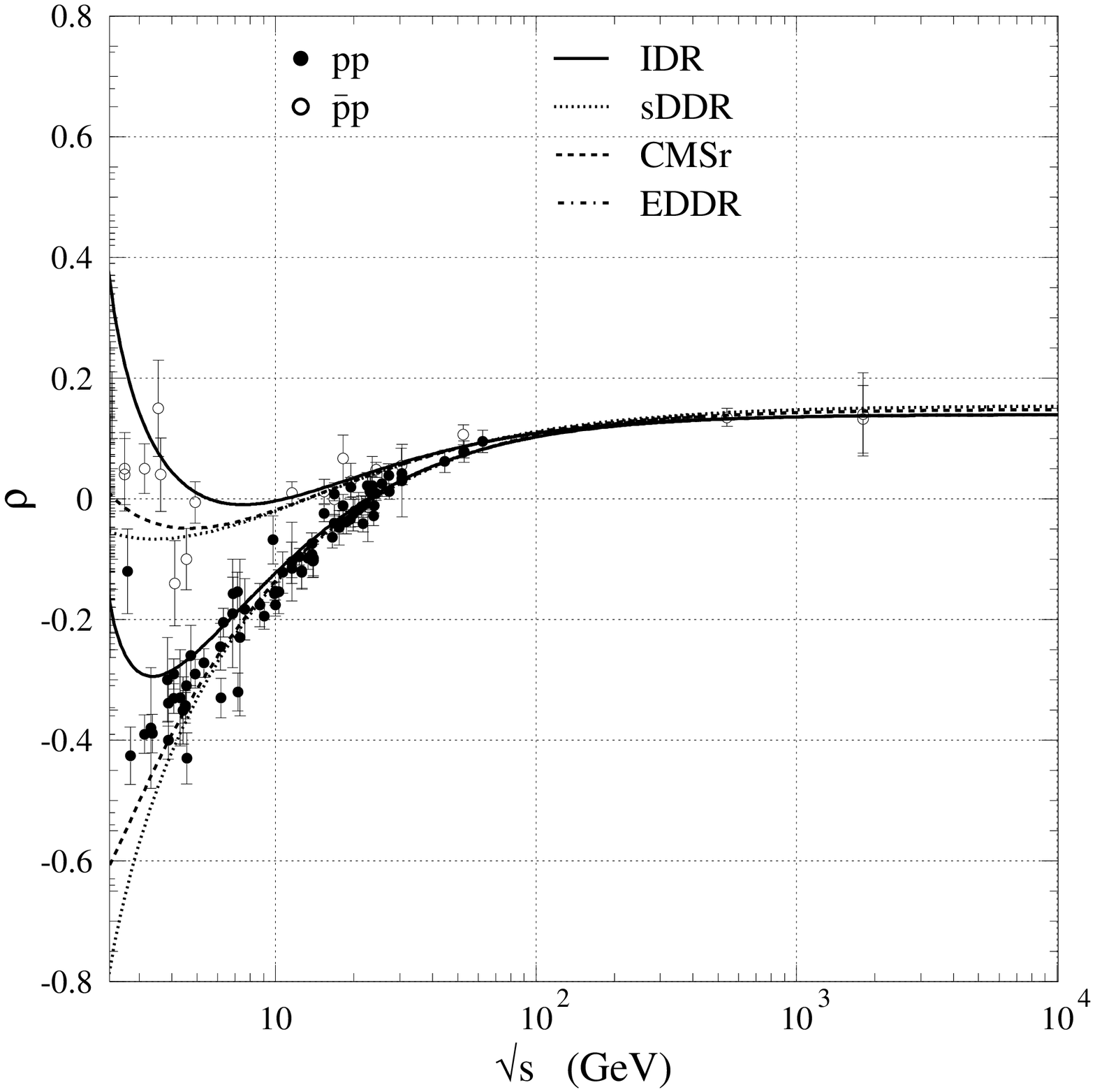}
\caption{Results of the simultaneous fit to $\sigma_{\mathrm{tot}}$ and
$\rho$ from $pp$ and $\bar{p}p$ scattering, by means of
Integral Dispersion Relations (IDR),
standard Derivative Dispersion Relations (sDDR), Cudell-Martynov-Selyugin
representation (CMSr) and the Extended Derivative Dispersion Relations (EDDR)
and considering the subtraction constant $K = 0$ (Table \ref{tab:1}). The curves corresponding
to IDR (solid) and EDDR (dot-dashed) coincide. } \label{fig:1}
\end{figure}

\begin{figure}[!]
\includegraphics[width=8.cm,height=6.cm]{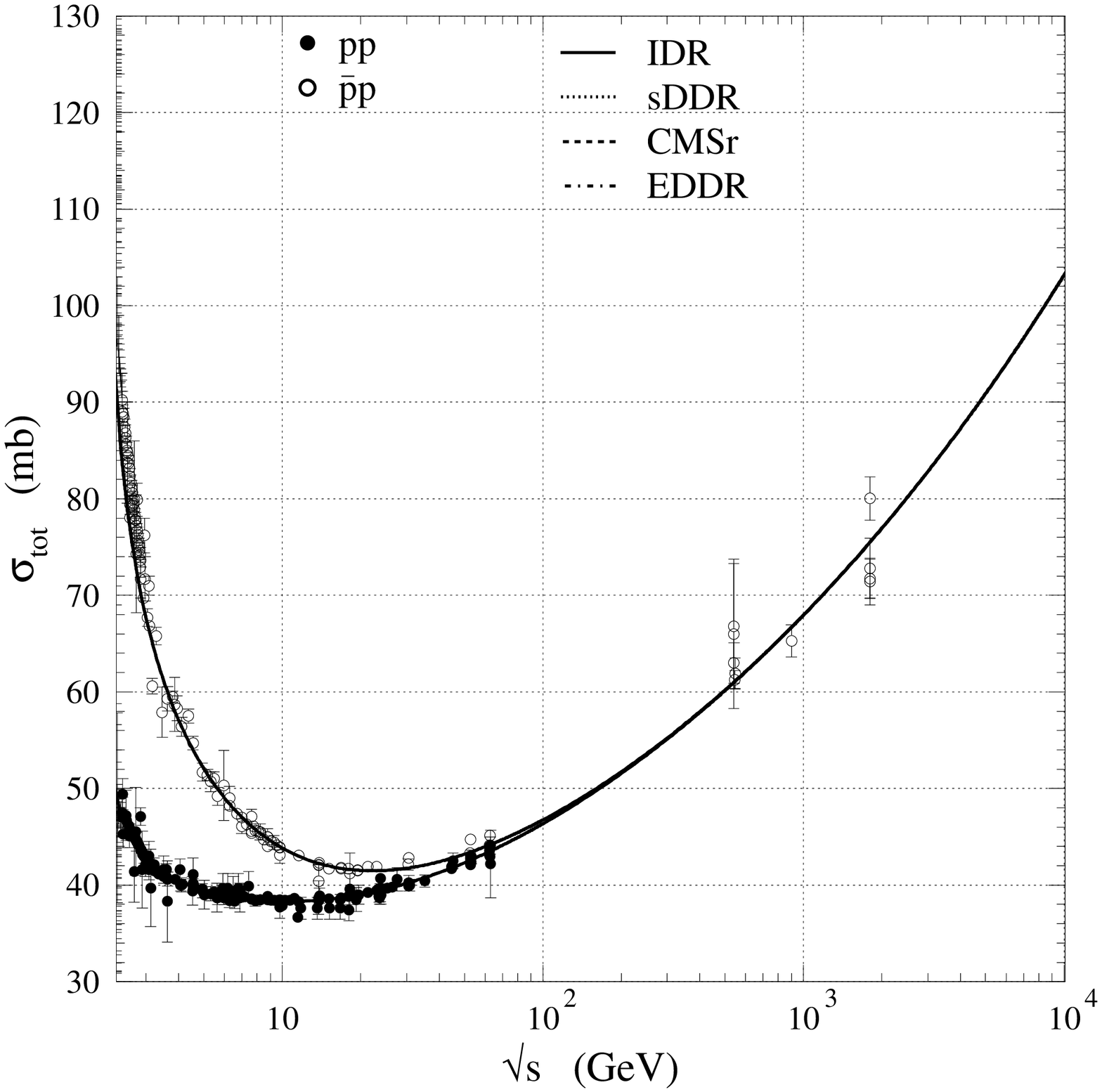}
\includegraphics[width=8.cm,height=6.cm]{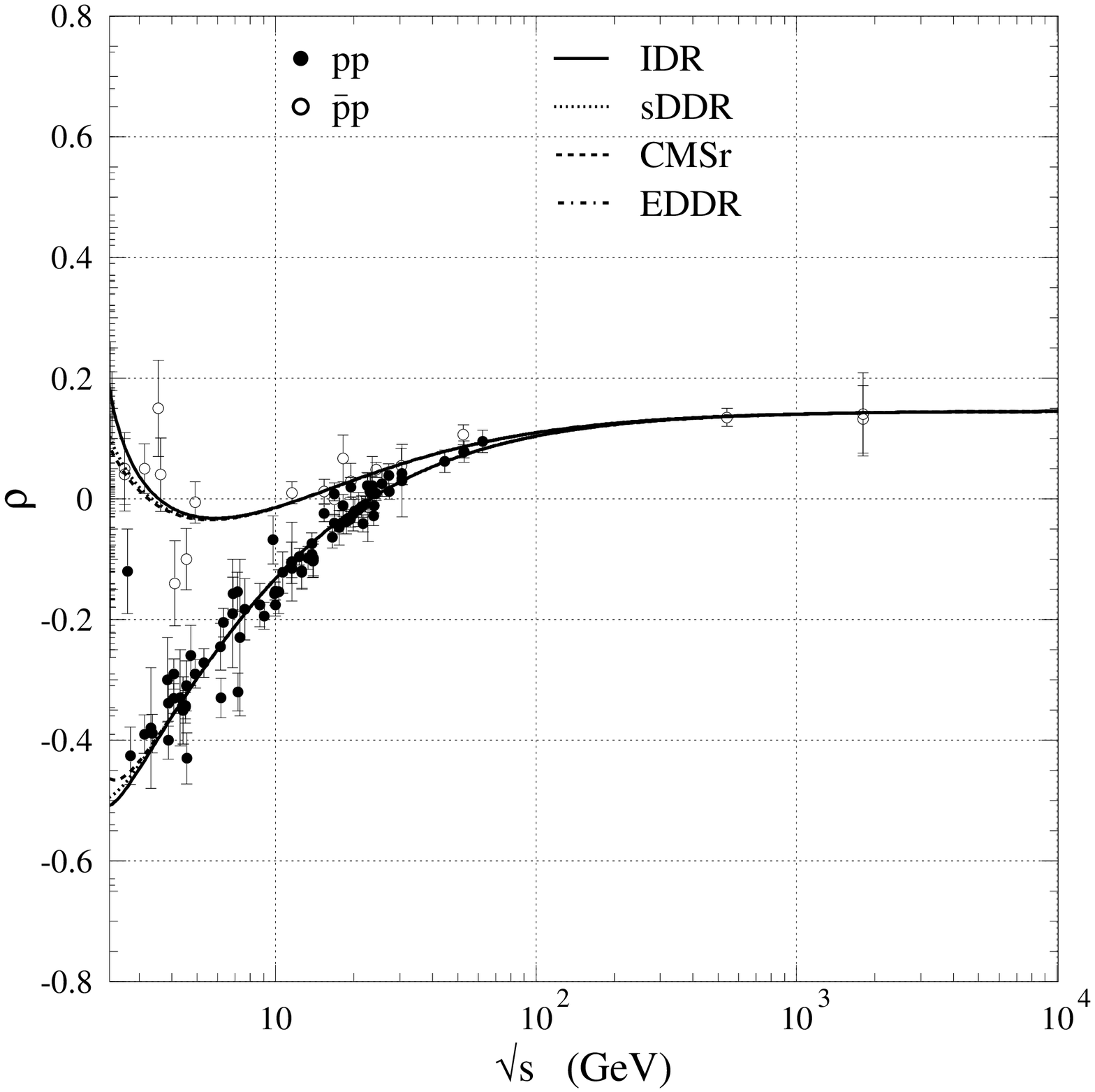}
\caption{Same as Figure \ref{fig:1} but considering the subtraction constant $K$ as
a free fit parameter (Table \ref{tab:2}).} \label{fig:2}
\end{figure}

In each of the four cases (IDR, sDDR, CMSr and EDDR), we consider two 
variants of the fits, one neglecting the subtraction constant (that 
is, taking $K = 0$) and the other considering the subtraction constant
as a free fit parameter. The numerical results and statistical 
information on the fits are displayed in Table \ref{tab:1} ($K=0$) and 
Table \ref{tab:2} ($K$ free). The corresponding curves together with 
the experimental data are shown in Fig. \ref{fig:1} ($K=0$) and Fig. 
\ref{fig:2} ($K$ free).

\begin{table*}[!]
\begin{center}
\caption{Simultaneous fits to $\sigma_{\mathrm{tot}}$ and $\rho$,
from $pp$ and $\bar{p}p$ scattering, for
$\sqrt s_{\mathrm{min}} =$ 4 GeV
(270 data points), with $K = 0$ and using Integral Dispersion Relations (IDR),
standard Derivative Dispersion Relations (sDDR), Cudell-Martynov-Selyugin
representations (CMSr) and the Extended Derivative Dispersion Relations (EDDR).}
\label{tab:1}
\begin{ruledtabular}
\begin{tabular}{ccccc}
& IDR & sDDR & CMSr & EDDR  \\
\hline
$X$ (mb)   & 1.662  $\pm$ 0.033  & 1.497  $\pm$ 0.032  & 1.563  $\pm$ 0.033  &1.662  $\pm$ 0.033 \\
$Y_+$ (mb) & 4.089  $\pm$ 0.058  & 3.800  $\pm$ 0.041  & 3.892  $\pm$ 0.047  &4.089  $\pm$ 0.058 \\
$Y_-$ (mb) &-2.143  $\pm$ 0.084  &-1.947  $\pm$ 0.070  &-2.039  $\pm$ 0.076  &-2.143 $\pm$ 0.084 \\
$\epsilon$ & 0.0884 $\pm$ 0.0021 & 0.0975 $\pm$ 0.0021 & 0.0939 $\pm$ 0.0021 &0.0884 $\pm$ 0.0020 \\
$\eta_+$   & 0.3797 $\pm$ 0.0099 & 0.3209 $\pm$ 0.0076 & 0.3427 $\pm$ 0.0087 &0.3797 $\pm$ 0.099 \\
$\eta_-$   & 0.569  $\pm$ 0.011  & 0.5583 $\pm$ 0.0098 & 0.567  $\pm$ 0.010  &0.569  $\pm$ 0.011 \\
$\chi^2$   & 382.1               & 365.4               & 325.9               &382.1 \\
$\chi^2/DOF$ & 1.45                & 1.38                & 1.23                & 1.45\\
\end{tabular}
\end{ruledtabular}
\end{center}
\end{table*}

\begin{table*}[!]
\begin{center}
\caption{Same as Table \ref{tab:1} but considering the subtraction constant $K$ as
a free fit parameter.}
\label{tab:2}
\begin{ruledtabular}
\begin{tabular}{ccccc}
& IDR & sDDR & CMSr & EDDR  \\
\hline
$X$ (mb)   & 1.598  $\pm$ 0.034  & 1.598  $\pm$ 0.034  & 1.598  $\pm$ 0.034  & 1.598  $\pm$ 0.034\\
$Y_+$ (mb) & 3.957  $\pm$ 0.053  & 3.957  $\pm$ 0.053  & 3.957  $\pm$ 0.053  & 3.957  $\pm$ 0.053\\
$Y_-$ (mb) &-2.082  $\pm$ 0.080  &-2.083  $\pm$ 0.079  &-2.084  $\pm$ 0.080  &-2.082  $\pm$ 0.080\\
$\epsilon$ & 0.0919 $\pm$ 0.0021 & 0.0919 $\pm$ 0.0021 & 0.0919 $\pm$ 0.0022 & 0.0919 $\pm$ 0.0021\\
$\eta_+$   & 0.3554 $\pm$ 0.0098 & 0.3555 $\pm$ 0.0097 & 0.3555 $\pm$ 0.0098 & 0.3554 $\pm$ 0.0098\\
$\eta_-$   & 0.569  $\pm$ 0.010  & 0.569  $\pm$ 0.010  & 0.569  $\pm$ 0.010  & 0.569  $\pm$ 0.010\\
$K$        &-2.27   $\pm$ 0.28   & 2.28   $\pm$ 0.33   & 1.00   $\pm$ 0.29   &-2.27   $\pm$ 0.28\\
$\chi^2$   & 315.4               & 314.6               & 314.2               & 315.4\\
$\chi^2/DOF$ & 1.20              & 1.20                & 1.19                &  1.20\\
\end{tabular}
\end{ruledtabular}
\end{center}
\end{table*}

\subsection{Discussion}

The main goal of this section is to discuss the consistences among
the results obtained by means of distinct analytical connections 
between the real and imaginary parts of the amplitude. However, some
phenomenological consequences can also be inferred from this study, as
discussed in what follows.

From Tables \ref{tab:1} and \ref{tab:2} we see that, as expected, the
best statistical results are obtained with the subtraction constant 
$K$ as a free fit parameter. However, as we shall show, taking $K=0$ 
gives suitable information not only on the practical equivalence
between the IDR and the differential forms (sDDR, CMSr and EDDR), but 
also on the important role played by the subtraction constant. For 
that reason we shall treat separately the cases  $K=0$ and $K$ as a 
fit parameter.

\subsubsection{Neglecting the subtraction constant}

From Table \ref{tab:1} we see that, for $K=0$, the numerical results
obtained with the IDR and the EDDR are exactly the same, up to four 
figures and that this does not occur in the case with the sDDR and the
CMSr neither. That is an important result since it demonstrates the 
accuracy of our analytical results for the extended derivative 
relations.

We note that the high values of $\chi^2/DOF$, in all the cases, are 
consequences of the specific analytical model considered (intended for 
the high-energy region) and the energy cutoff used. We add the fact 
that we did not performed any data selection, but used all the 
available data	from the PDG archives. However, as already commented
this disadvantage has no influence in our main goal, namely tests of
consistences.

The effects of the equivalences (IDR and EDDR) and differences (IDR
and sDDR or CMSr) in the description of the experimental data are 
shown in Fig. \ref{fig:1}. The curves corresponding to IDR (solid) and 
EDDR (dot-dashed) coincide at all the energies above the threshold and 
we see that even with the fit cutoff at 
$\sqrt s_{\mathrm{min}} = 4$ GeV, the description of the experimental 
data below this point is reasonably good in both cases. On the other 
hand, the differences between the exact results (IDR and EDDR) and the 
sDDR or CMSr are remarkable for $\sigma_{\mathrm{tot}}(s)$ at the
highest energies and for $\rho(s)$ in the region of low energies 
(below $\sqrt s \approx 10$ GeV).

In the case of the total cross section, the results with sDDR and CMSr
indicate a faster increase with the energy then those with the IDR and 
EDDR. We stress the importance of this point, since it gives different 
solutions for the well known puzzle between the CDF data \cite{cdf} 
and the E710/E811 data \cite{e710,e811} at $\sqrt s = 1.8$ TeV; in 
this respect, we see that the exact results (IDR and EDDR) favor the 
E811/E710 results. In particular the values for the Pomeron intercept 
read (Table \ref{tab:1}): $\alpha_{\tt I\!P}(0) = 1.0884 \pm 0.0021$ 
(IDR and EDDR), $1.0975 \pm 0.0021$ (sDDR) and $1.0939 \pm 0.0021$ 
(CMSr).

\subsubsection{Subtraction constant K as a free fit parameter}

With $K$ as a free fit parameter our results demonstrate, once more, 
an effect that we have already  noted before \cite{am04}, namely the 
high-energy approximation can be absorbed by the subtraction constant.
In fact, from Fig. \ref{fig:2} we see that in this case, the 
differences between the sDDR/CMSr and the exact results IDR/EDDR,
practically disappear. From Table \ref{tab:2} we can identify the 
subtraction constant as the responsible for this complementary effect:
the numerical values of the fit parameters and errors are practically
the same in all the four cases, except for the values of $K$, that is,
in practice, the differences are absorbed by this parameter. We 
conclude that the subtraction constant affects the fit results even in
the region of the highest energies; this effect is due to the
correlations among the free parameter in the fit procedure, as
previously observed \cite{am04,alm03}. Of course, also in this case 
the numerical values obtained with the IDR and EDDR are exactly the 
same, including the value of the subtraction constant up to four 
figures (Table \ref{tab:2}).

In particular, we note that all the four variants indicate the same 
result for the intercept of the Pomeron, namely 
$\alpha_{\tt I\!P}(0) = 1.0919 \pm 0.0021$. The corresponding result 
for the total cross section lies nearly between the CDF and E811/E710 
results, barely favoring the last ones (Figure \ref{fig:2}).

\section{Critical Remarks \label{sec:critical}}

We have demonstrated that for the class of functions defined by
Theorem \ref{theo:1}, IDR can be formally replaced by differential 
operators without any high-energy approximation; we have also verified 
the equivalence between the integral and extended derivative results, 
in the particular case of a simple phenomenological parametrization
for $pp$ and $\bar{p}p$ scattering. Despite the encouraging results
reached the whole analysis has limitations from both formal and
practical points of view. In what follows we summarize the main
critical points, giving references where more details can be found and 
providing also suggestions for further investigations. Since the EDDR 
involve two contributions, the tangent operators (sDDR) and the 
correction terms $\Delta^{+/-}(E, m)$, we shall consider the two cases
separately. We also present some critical remarks on the
Pomeron-Reggeon model used in Sect. \ref{sec:pra} as a practical 
framework.

\subsection{sDDR}

First, let us discuss some aspects related to the dispersion approach
as it has been treated and widely used in the literature till now,
namely the sDDR, Eqs. (\ref{eq:sddre}) and (\ref{eq:sddro}). As 
commented in Sect. \ref{sec:SDDR}, these equations are obtained
by considering the limit $m \rightarrow 0$ in the IDR,
Eqs. (\ref{eq:idre}) and (\ref{eq:idro}).
This condition is a critical one, which puts serious practical
and formal limitations in any use of the sDDR, because that means to 
go to lower energies by passing through different thresholds, 
resonances, poles, up to $E = 0$! In this sense, the expression of the
tangent operator as an integral from $E = 0$ to $E = \infty$ does not 
guarantee any local character for the differential approach (or the 
corresponding integral), even in the case of convergence of the 
series. In other words, this representation of the non-local operator
(integral) in terms of local operators (tangent series), does not 
guarantee the non-locality of the result. Moreover, the representation
does not apply near the resonances \cite{ed} and the convergence of 
the series has been discussed in several works leading some authors to
argue that, in a general sense, the mathematical condition for the 
convergence ``excludes all cases of physical interest" \cite{ed}. 
These and other aspects were extensively discussed in the seventies 
and eighties \cite{ed,hk,hjk,bd,ddr-others,kf,kf-blois} and some
points have been recently reviewed in \cite{am04}.

However, there is a fundamental point developed by some authors that
enlarger the practical applicability of the sDDR under some special
conditions \cite{ed,kf}. As stated by Kol\'a\v{r} and Fischer 
\cite{kf-blois}, in discussing the replacement of IDR by sDDR we must
distinguish two formulations: (1) to consider the case of asymptotic 
energies and a finite number of terms in the tangent series; (2) to 
consider finite energies and an infinity number of terms in the series.
The former case applies for smooth behaviors of the amplitude (as it
is the case at sufficiently high energies, specifically 
$\sqrt s >$ 10 - 20 GeV).
That includes a wide class of functions of physical interest, mainly
if only the first term can be considered \cite{ed,kf} (see \cite{acmm06}
for a recent analysis even beyond the forward direction).
The later case, however, is critical for at least two reasons. First,
because the condition of convergence of the series 
(Theorem \ref{theo:1}) limits the class of functions of practical 
applicability. Secondly and more importantly, since the high-energy 
approximation is enclosed, all the strong limitations referred to 
before applies equally well to this case. In conclusion, the class of 
functions for which the sDDR have a practical applicability depends 
strongly on the formalism considered and is narrower in the case of 
finite energies, namely entire functions in the logarithm of the energy.

\subsection{EDDR}

Let us now discuss the EDDR, with focus on the role of the correction
terms $\Delta^{+/-}(E, m)$ in Eqs. (\ref{eq:eddre}) and 
(\ref{eq:eddro}). First we note that these infinity series are 
analytically associated with the fixed lower limit $m$ in the integral
representation and they correspond to the contributions that are 
neglected in the high-energy approximation ($m \rightarrow 0$). For 
example, for the even case we have the formal identity
\begin{eqnarray}
\tan\left(\frac{\pi}{2}
\frac{{\mathrm{d}}}{{\mathrm{d}}\ln E}\right)\frac{\Im F_+(E)}{E}
+ \frac{\Delta^+(E,m)}{E}
= \nonumber \\
=\frac{2E}{\pi}P\!\!\!\int_{m}^{+\infty}
\!\!\!\d E'
\frac{1}{E'(E'^{2}-E^{2})}\Im F_{+}(E'),
\nonumber
\end{eqnarray}
which means that all the physical situation concerns the region
\textit{above} the physical threshold $E = m$. Therefore, from a formal
point of view, the critical points raised above on the sDDR
(tangent operator only), concerning the infinity series in the region 
$E:0 \rightarrow m$, do not apply in this case and the critical point 
here concerns only the convergence of the correction series and their 
practical applicability.

On the one hand, from a formal point of view (as already discussed at 
the end of Sect. \ref{sec:EDDR}), the convergence of the correction 
series is ensured by Theorem \ref{theo:1} and that means a narrower 
class of functions than that associated only with entire functions in 
the logarithm of the energy. This restriction is due to the infinity 
number of derivatives in $\ln s$.
We shall give and discuss some examples in what follows.

From a practical point of view, it is obvious that the efficiency 
and/or real applicability of not only the derivative approach (EDDR),
but also the integral one (IDR), depends on the specific physical 
problem involved. In this scenario (the physical problem) we expect to
find some specific limitations that are independent of the formal 
aspects referred to above and these aspects demands also some comments.
In principle and in a \textit{general sense}, if we attempt to apply
dispersion techniques directly to the experimental data (related to 
the imaginary part of a function), we are faced with the problem of
error propagation from the experimental uncertainties. Even if we can
``reproduce" the experimental behavior by means of suitable analytical 
parameterizations, with statistical errors inferred for the free 
parameters, these errors should, in principle, be propagated too. In 
this case, the infinity series in both sDDR and EDDR have certainly 
limited usefulness (see for example \cite{hjk,bd} for the sDDR case).
However, if error propagation from the fit results is not of interest 
or can be neglected, and, most importantly, one has a ``correct" or 
acceptable model for the imaginary part of the amplitudes, then we are 
restricted only to the the formal conditions discussed above and the 
derivative approach becomes reliable, including the correction series 
(Theorem \ref{theo:1}).

Let us now discuss the specific physical problem that motivated the
present analysis. As commented in our introduction and in Sect. 
\ref{sec:DR}, we focused the dispersion techniques in the context of 
hadron scattering, in special in the elastic case, for which a complet 
theory is still absent. The main goal concerns the connections between 
total cross section and the $\rho$ parameter for energies above 
$\sqrt s$ = 5 - 10 GeV.  In terms of dispersion techniques the usual 
way to treat the subject is by means of IDR, sDDR and the analyticity 
prescriptions for even and odd amplitudes (and recently the CMSr).
In this specific case, besides the absence of a pure QCD treatment, 
the subject is characterized by three kinds of problems: 
(1) formal justification of the usual phenomenology;
(2) approximated descriptions of the experimental data by 
phenomenological models;
(3) experimental data available (problems (2) and (3) are certainly 
connected). Since these problems affect the practical applicability 
and efficiency of the dispersion techniques, let us shortly discuss 
some aspects involved.

(1) As it is well know, the usual phenomenology for the total cross 
sections is based on the reggeon concepts and involves distinct 
contributions from Pomerons and secondaries Reggeons. In this context,
analytical parameterizations for the total cross sections are 
characterized by power and logarithm functions of the energy 
(Reggeons, simple, double and triple pole Pomerons) and the fits are 
performed not below  $\sqrt s$ = 5 GeV. We note that all these 
contributions belong to the class of functions defined by Theorem 
\ref{theo:1} (the tangent series can be summed leading to closed 
analytical results) and they have been used and investigated in several
works \cite{cms,ckk,alm03,compete}. However, as we have already
pointed out \cite{am04}, the central problem here concerns the fact 
that these contributions are formally justified only for asymptotic 
energies ($E$ or $s$ $\rightarrow \infty$), which certainly is not the
case for the energies considered. The applicability of these models 
seems to be justified only under the hypothesis that the accelerators 
have already reached the energies that can be considered asymptotic in 
the mathematical context, which seems to us a dangerous assumption.

(2 - 3) A close look at the bulk of experimental data available shows 
that these ensembles present several discrepancies due to spurious 
data, normalization problems and other effects. In this respect, 
recent analysis have pointed out the necessity of some screening 
criterion in order to select the ``correct" experimental information.
We shall not discuss this question here because it seems to us an open
problem. But the point is that this fact puts serious limitations in 
any interpretation of statistical tests of the fits, as the popular 
$\chi^2$ per degree of freedom and, consequently, not only in the
efficiency of the phenomenological descriptions, but also
in the possible selection of the best phenomenological model.

At last let us return to the applicability of the EDDR, now in this 
context. Despite of all the above problems, the known and usual 
phenomenological approach is characterized by analytical 
parameterizations for the imaginary parts of the amplitude and 
statistical tests on the quality of the fit. In this case, with 
specific analytical representations for the total cross sections, 
without error propagation from the fit parameters and in the Regge 
context we understand that the correction terms, we have introduced, 
can have a suitable applicability in the context of the dispersion
techniques. The point is that the class of functions for which they 
hold includes all the usual Regge parameterizations and since the 
high-energy approximations is absent the fits can be formally extend 
to lower energies. However to reach a good statistical description of 
the data, specifically near the threshold, demands a ``correct" model
for the imaginary part of the amplitude, which, to our knowledge is 
still lacking. We shall return to this point in what follows.

\subsection{Pomeron-reggeon parametrization}

Based on all the above limitations in the phenomenological context, we
have chosen one of the possible (and popular) models in order to check
the equivalences (and differences) among the different dispersion 
representations analyzed in this work. Although, as demonstrated in 
Sect. \ref{sec:pra}, this choice is sufficient for our aim, some 
drawnbacks involved demand additional comments.

In the mathematical context, as demonstrated by Kol\'a\v{r} and 
Fischer \cite{kf}, some formal results, theorems and representations
for the derivative relations were obtained under the assumption of the 
Froissart-Martin bound, $\sigma_{\mathrm{tot}} < c\ln^2 s$, but other 
forms of sDDR do not require this bound. Therefore, since the simple 
pole Pomeron contribution, that dominates at the asymptotic energies,
violates this bound, the model assumed is not an example in full 
agreement with the totality of the formal results. However, as already 
exemplified (Sect. \ref{sec:SDDR}), the model belongs to the class of 
functions defined by Theorem \ref{theo:1} and therefore, in this 
restrictive sense it seems to us to be an acceptable choice.

In the formal phenomenological context, when applied below asymptotic
(infinity) energies, the model suffers from all the drawbacks already 
discussed. Despite of this, its use above, let us say, 
$\sqrt s$ = 10 GeV, could be explained (not justified) by the fact 
that the Regge approach is the only known formalism, able to describe 
some global characteristics of the soft scattering. What is presently 
expected is the development of a microscopic theory able to justify 
its efficiency.

Now let us focus in the low energy region, above the physical 
threshold, $2m_p \approx$ 1.88 GeV $< \sqrt s \leq$ 10 GeV and discuss
the usefullness and practical applicability of the EDDR. To our 
knowledge, there is no model proposed for this interval and that could
explain the fact that fit procedures, even through IDR, make use of 
energy cutoffs at $\sqrt s \approx$ 5 GeV (\cite{ua4} is a typical 
example). In this sense, the usefulness of the correction terms 
$\Delta^{+/-}(E, m)$ could be questioned. However, we understand that
the lack of a phenomenological approach for that region may also be a 
mirror of the present stage, characterized by a focus (probably 
excessive) on the highest and asymptotic energies (the great 
expectations from the Tevatron, RHIC, LHC). In our oppinion, 
independently of the fact that the ``asymptopia" might be resolved or 
not in a short term, the connection between resonance region (above 
the physical threshold) and the high-energy region (above 10 GeV) 
still remains a fundamental problem demanding  solution. In this 
respect we understand that the EDDR can play an important role in 
further investigations.

Concerning the practical applicability of the extended relations in 
this region, it is obviously limited, due to the lack of a ``correct" 
or accepted analytical model for the imaginary part of the amplitude. 
One way to circumvent this problem could be the introduction
of a different parametrization for this particular region. That was 
the procedure used in Ref. \cite{cms}; although without justification 
or explicit reference to the analytical form used, the authors
obtained reasonable fit results. However, beyond the lack of any 
physical meaning, this procedure puts limitations on the equivalence
between integral and derivative representations.

Based on the above facts and aimed only to check and compare the 
results obtained through different dispersion representations, we
considered the Pomeron-Reggeon parametrization extended up to the low
energy region, with a fit cutoff at $\sqrt s$ = 4 GeV. Certainly the 
statistical results displayed in Tables \ref{tab:1} and \ref{tab:2} 
indicate that the confidence level is very low and even a look at 
Figs. \ref{fig:1} and \ref{fig:2} shows that the data near the 
resonance are not adequately described. As a consequence the 
numerical results in Tables \ref{tab:1} and \ref{tab:2} may be 
questionable on physical grounds. However, we insist that all the 
figures in these Tables are fundamental for a definite check of all 
the analytical representations investigated (which is the only aim of
Sect. \ref{sec:pra}). At last we note that one may think that it might
be possible to find a suitable function, in agreement with the 
convergence condition and able to fit all the experimental data of 
interest on secure statistical grounds; that would be enough for our 
tests of consistences. We are not sure about this possibility, but
the point is that the use of a known and popular parametrization, 
even with limited efficiency, can bring new insights for further 
developments mainly because it gives information
on what should be improved.

\section{Conclusions and final remarks} \label{sec:conclusions}

We have obtained novel analytical expressions for the derivative
dispersion relations, without high-energy approximations. The
mathematical results
are valid for the class of functions specified by
Theorem \ref{theo:1}.
In principle, their applicability can be extended to any area
that makes use of dispersion techniques, with possible additional
constraints, dictated by the analytical and experimental conditions
involved. In special, under aadequate circumstances, the local 
character of the derivative operators may be a great advantage.

For scattering amplitudes belonging to the class of functions
defined by Theorem \ref{theo:1}, the EDDR are valid for
any energy above the physical threshold.
Since the experimental data on the total cross sections
indicate a smooth variation with the energy (without oscillations
just above the physical threshold and a smooth systematic increase 
above $\sqrt s \approx 20$ GeV), this class includes the majority of
functions of physical interest. Using as framework a popular 
Pomeron-Reggeon parametrization for the total cross sections,
we have checked the numerical equivalence between the results obtained
with the IDR (finite lower limit $m$) and the EDDR, as well as the 
differences associated with the sDDR and the CMSr. We have also 
presented a critical discussion on the limitations of the whole 
analysis from both formal and practical points of view.

We stress that, as in the case of IDR, the practical
efficiency of the EDDR in the reproduction of the experimental data on
$\sigma_{\mathrm{tot}}$ and $\rho$ depends on the model considered. 
Here, in order only to check the consistences among the different
analytical forms, we made use of a particular Pomeron-Reggeon 
parametrization, for which a cutoff at $\sqrt s =$ 4 GeV was 
necessary. For example, by considering
the full nondegenerated case (four contributions, each one from each
meson trajectory, $a_2, f_2, \rho, \omega$), this cutoff can be reduced
\cite{digi}, or the $\chi^2/DOF$ can be reduced for the same cutoff.
Despite the limitations of our practical example 
(Sec. \ref{sec:critical}), some interesting phenomenological aspects 
could be inferred. In particular, although already noted 
\cite{am04,alm03}, we have called the attention to the role of the 
subtraction constant as a practical ``regulator", in the replacement 
of IDR by derivative forms, a fact that is clearly identified in 
Table \ref{tab:2}: the high-energy approximation is absorbed by the
constant. In this respect, we have demonstrated that this artifice, 
which lack physical meaning, can be avoided by the direct use of the 
EDDR. However, this observation does not depreciate the important role 
of the subtraction constant as a free fit parameter, since the best 
statistical results are obtained in this context (Tables \ref{tab:1} 
and \ref{tab:2}). In particular, we note that the effect of this 
parameter is to provide a slight higher value for the Pomeron 
intercept, $\alpha_{\tt I\!P}(0) \approx 1.088$ ($K=0$) and
$\alpha_{\tt I\!P}(0) \approx 1.092$ ($K$ free).

To our knowledge, a well established theoretical approach for total
cross sections just above the physical threshold and in the region
connecting low and high energies is still absent. In this sense,
despite all the limitations discussed, we hope that the local analytical
operators, developed here for these regions, can contribute, as a
formal mathematical tool, for further developments on the subject.

\begin{acknowledgments}

We are thankful to FAPESP for financial support
(Contracts No.03/00228-0 and No.04/10619-9).

\end{acknowledgments}

\end{document}